\documentclass[aps,prl,epsf,twocolumn,showpacs,groupedaddress,nofootinbib]{revtex4}
\usepackage{graphicx}
\usepackage{amsmath,amssymb,latexsym}
\usepackage{bm}
\newcommand{\lsim}{\mathrel{\vcenter{\hbox{$<$}\nointerlineskip\hbox{$\sim$}}}}
\newcommand{\gsim}{\mathrel{\vcenter{\hbox{$>$}\nointerlineskip\hbox{$\sim$}}}}

\newcommand{\eps}{\epsilon}

\newcommand{\En}{E_n}
\newcommand{\Fn}{F_n}
\newcommand{\Enu}{E_{\overline{\nu}}}

\newcommand{\Enubar}{\epsilon_{\overline{\nu}}}
\newcommand{\thnu}{\overline{\theta}_{\overline{\nu}}}
\newcommand{\cth}{\cos\overline{\theta}_{\overline{\nu}}}
\newcommand{\Fnu}{F_{\overline{\nu}}}
\newcommand{\tbar}{\overline{\tau}_n}
\newcommand{\Enmin}{E_n^{\rm min}}
\newcommand{\Enmax}{E_n^{\rm max}}
\newcommand{\epsmin}{\epsilon_{\overline{\nu}}^{\rm min}}
\newcommand{\nuebar}{\overline{\nu}_e}

\newcommand{\bwide}{\begin{widetext}}
\newcommand{\ewide}{\end{widetext}}
\newcommand{\beq}[1]{\begin{equation} \label{(#1)}}
\newcommand{\eeq}{\end{equation}}
\newcommand{\ba}[1]{\begin{eqnarray} \label{(#1)}}
\newcommand{\ea}{\end{eqnarray}}

\newcommand{\rf}[1]{(\ref{(#1)})}
\begin{document}
\title{Galactic Point Sources of TeV Antineutrinos}
\author{Luis A. Anchordoqui$^a$, Haim Goldberg$^a$, Francis Halzen$^b$, and
Thomas J. Weiler$^{b,c,d}$}
\affiliation{$^a$Department of Physics, Northeastern University, Boston MA 02155\\
$^b$Department of Physics, University of Wisconsin, Madison WI 53706\\
$^c$Center for Cosmological Physics, University of Chicago, Chicago IL 60637\\
$^d$Department of Physics and Astronomy, Vanderbilt University, Nashville TN 37235
}
\begin{abstract}
High energy cosmic ray experiments have identified an
excess from the region of the
Galactic Plane in a limited energy range around $10^{18}$~eV~(EeV).
This is very suggestive of neutrons as candidate primaries, because the
directional signal requires relatively-stable neutral primaries,
and time-dilated neutrons can reach Earth from typical Galactic distances
when the neutron energy exceeds an EeV.
We here point out that if the Galactic messengers are neutrons,
then those with energies below an EeV
will decay in flight, providing a flux of cosmic
antineutrinos above a TeV which is {\it observable}
at a kilometer-scale neutrino observatory.
The expected event rate per year above 1 TeV in a detector such as IceCube,
for example,
is 20 antineutrino showers (all flavors) and
a $1^\circ$ directional signal of 4 $\bar \nu_\mu$ events.
A measurement of this flux can serve to identify
the first extraterrestrial point source of TeV antineutrinos.
\end{abstract}


\maketitle

An intriguing anisotropy in the cosmic ray spectrum has emerged
in the energy range near an EeV. The Akeno Giant Air Shower Array
(AGASA) has revealed a correlation of the arrival direction of the cosmic
rays to the Galactic Plane (GP) at the $4\sigma$
level~\cite{Hayashida:1998qb}. The GP excess, which is roughly 4\% of
the diffuse flux, is mostly
concentrated in the direction of the Cygnus region, with a second
spot towards the Galactic Center (GC)~\cite{Teshimaicrc}. Evidence at
the 3.2$\sigma$ level for GP
enhancement in a similar energy range has also been reported by the Fly's Eye
Collaboration~\cite{Bird:1998nu}. The existence of a point-like excess in the
direction of the GC has been confirmed via independent
analysis~\cite{Bellido:2000tr} of data collected with the Sydney University
Giant Airshower Recorder (SUGAR). This is a remarkable level of
agreement among experiment using a variety of techniques.

Independent evidence may be emerging for a cosmic
accelerator in the Cygnus spiral arm. The HEGRA experiment has detected an
extended TeV $\gamma$-ray source in the Cygnus region with no clear
counterpart and a spectrum not easily accommodated with synchrotron radiation
by electrons~\cite{Aharonian:2002ij}. The difficulty to accommodate the
spectrum by conventional electromagnetic mechanisms has been exacerbated
by the failure of CHANDRA and VLA to detect X-rays or radiowaves signaling
acceleration of any electrons~\cite{Butt:2003xc}. The model proposed is that
of a proton beam, accelerated by a nearby mini-quasar or possibly Cygnus X-3,
interacting with a molecular cloud to produce pions that are the source of
the gamma rays. Especially intriguing is the possible association of this
source with Cygnus-OB2, a cluster of more than 2700 (identified) young,
hot stars with a total mass of $\sim 10^4$ solar
masses~\cite{Knodlseder:2000vq}. Proton
acceleration to explain the TeV photon signal requires only 0.1\% efficiency
for the conversion of the energy in the stellar wind into cosmic ray
acceleration. Also, the stars in Cygnus-OB2 could be the origin of
time-correlated, clustered supernova remnants forming a source of cosmic
rays. By cooperative acceleration their energies may even
exceed the ${\sim}1$\,PeV cutoff of individual remnants and accommodate
cosmic rays up to the ankle, where the steeply
falling ($\propto E^{-3.16\pm 0.08}$) cosmic
ray spectrum flattens to $E^{-2.8 \pm 0.3}$~\cite{Nagano:1991jz}.

All evidence points to a transition from galactic to extragalactic sources
above several EeV of primary energy. The steepness of the falloff between
the knee (about 3 PeV) and the ankle (about 10 EeV) is expected from
supernova shock models,
and may indicate that we are witnessing the high energy end of
the galactic flux. The extension of the nominal PeV cutoff beyond the
ankle can be understood as a collective effect of
stellar winds originating in the region of multiple supernova explosions.
These provide a
second acceleration to the particles and boost their energies
far beyond the values expected from their single-shock encounter.
Strong additional support for this picture emerges from
accumulating evidence (in a bi-modal proton-iron model) for a
dominant Fe component in the flux~\cite{Dova:2003ng}. The
importance of the heavy component is apparent all the way down to
the region of several PeV~\cite{Swordy:2002df}, with the spectral
index hardening slightly to $3.02\pm 0.03$ below 500 PeV. {\sl An
immediate consequence of this nucleus-dominance picture is the
creation of free neutrons via nuclei photodisintegration on
background photon fields.} These liberated neutrons are presumably
responsible for the observed directional signals. {\sl This
implies that it may not be a coincidence that the signal appears
first at energies where the neutron lifetime allows propagation
distances of galactic scales, i.e., 10 kpc.}

For every surviving neutron at $\sim$~EeV,
there are many neutrons at lower energy that decay via
$n\rightarrow p+e^- +\nuebar.$
The decay mfp of a neutron is
$c\,\Gamma_n\,\tbar=10\,(E_n/{\rm EeV})$~kpc,
the lifetime being boosted from its rest-frame value
$\tbar=886$~seconds to its lab value via
$\Gamma_n=\En/m_n$. The proton is bent by the Galactic magnetic
field, the electron quickly loses energy via synchrotron radiation,
and the $\nuebar$ travels
along the initial neutron direction, producing a directed TeV energy
beam whose flux is calculable.
{\sl We show in this Letter that
the expected $\bar \nu$ flux from the direction of the Cygnus region
is measurable in IceCube}~\cite{icecube}.
Furthermore, by the same logic, a km--scale Mediterranean detector,
if designed with sufficiently low threshold, can
see the $\bar \nu$ flux pointing toward the GC source as well.
The GZK neutrinos~\cite{Greisen:1966jv} and
the ``essentially guaranteed'' $\bar \nu$ flux calculated here probably
constitute the best motivated cosmic neutrino fluxes.
Of these two neutrino fluxes, the expected event rate for the
galactic beam is higher: 4 $\bar \nu_\mu$ events per year
and 16 in $\bar \nu_e + \bar \nu_\tau$ showers.

We turn to the calculation.
The basic formula that relates the neutron flux at the source
($d\Fn/d\En$) to the antineutrino flux observed at Earth ($d\Fnu/d\Enu$) is:
\bwide
\beq{nuflux}
\frac{d\Fnu}{d\Enu}(\Enu)=
\int d\En\,\frac{d\Fn}{d\En}(\En)
\left(1-e^{-\frac{D\,m_n}{\En\,\tbar}}\right)\,
\int_0^Q d\Enubar\,\frac{dP}{d\Enubar}(\Enubar)
\int_{-1}^1 \frac{d\cth}{2}
\;\delta\left[\Enu-\En\,\Enubar\,(1+\cth)/m_n\right]
\,.
\eeq
\ewide
The variables appearing in Eq.~\rf{nuflux} are the antineutrino and
neutron energies in the lab ($\Enu$ and $\En$),
the antineutrino angle with respect to the direction of the
neutron  momentum, in the neutron rest-frame ($\thnu$),
and the antineutrino energy in the neutron rest-frame
($\Enubar$).  The last three variables are not observed
by a laboratory neutrino-detector, and so are integrated over.
The observable $\Enu$ is held fixed.
The delta-function relates the neutrino energy in the lab to the
three integration variables.
The parameters appearing in Eq.~\rf{nuflux} are the
neutron mass and rest-frame lifetime ($m_n$ and $\tbar$),
and the distance to the neutron source ($D$).
$dF_n/dE_n$ is the neutron flux at the source,
or equivalently, the neutron flux that would be observed
from the source region in the absence of neutron decay.
Finally, $\frac{dP}{d\Enubar}(\Enubar)$ is the
normalized probability that the
decaying neutron in its rest-frame produces a $\nuebar$ with
energy $\Enubar$;
$\int^Q_0 d\Enubar\;\frac{dP(\Enubar)}{d\Enubar}=1$
defines the normalization, where the maximum neutrino energy in the
neutron rest frame is just 
$Q\equiv m_n -m_p -m_e = 0.71$~MeV,
and the minimum neutrino energy is zero in the massless
limit.\footnote{The massless-neutrino approximation seems justifiable here:
even an eV-mass neutrino produced at rest in the neutron
rest-frame would have a lab energy of $m_\nu\,\Gamma_n\lsim$~GeV,
below threshold for neutrino telescopes.} For the decay of unpolarized neutrons,
there is no angular dependence in $\frac{dP}{d\Enubar}$.

The $\En$ integration in Eq.~\rf{nuflux}
is effectively cut off at $\sim$~EeV,
the energy beyond which a neutron is stable over a 10~kpc path-length,
and it is truly cut off by the $E_{\rm max}$ of the neutron spectrum.
The expression in parentheses in Eq.~\rf{nuflux} is the decay
probability for a neutron
with lab energy $\En$, traveling a distance $D$.
In principle, one should consider a source distribution,
and integrate over the volume $\int\,d^3 D$.
Instead, we will take $D$ to be the 1.7 kpc distance
from Earth to Cygnus OB2;
for the purpose of generating the associated neutrino flux,
this cannot be in error by too much.

The Galactic anisotropy observed by the various collaborations
spans the energy range 0.8 to 2.0 EeV.\footnote{Actually, the anisotropy reported
in~\cite{Bird:1998nu} peaks in the energy bin $0.4-1.0$ EeV, but
persists with statistical significance to energies as low as 0.2 EeV.
The full Fly's Eye data include a directional signal which was somewhat
lost in unsuccessful attempts~\cite{Cassiday:kw} to relate it to $\gamma$-ray emission from
Cygnus X-3. This implies that if neutrons are the carriers of the anisotropy,
there {\em needs to be} some contribution from at least one source
closer than 3-4 kpc.}
The lower cutoff specifies
that only neutrons with EeV energies and above have a boosted
$c\tau_n$ sufficiently large to serve as Galactic messengers.
The upper cutoff reflects an important feature
of  photodisintegration at the source: heavy nuclei with energies
in the vicinity of the ankle will fragment to neutrons with
energies about an order of magnitude smaller. To account for the
largest neutron
energies, it may be necessary to populate the heavier nucleus
spectrum in the region above the ankle. This is not a problem --
one fully expects the emerging harder extragalactic spectrum to
overtake and hide the steeply falling galactic population. It is
not therefore surprising that in order to fit the spectrum in the
anisotropy region and maintain continuity to the ankle region
without introducing a cutoff, the AGASA Collaboration required a
spectrum $\propto E^{-3}$ or steeper~\cite{Hayashida:1998qb}.

A detailed scenario for ultrahigh energy nuclei (parents of the
anisotropy neutrons) originating in a pulsar close to the Cygnus
OB2 region has been recently described~\cite{Bednarek:2003cx}.
The sequence begins with the one-shot acceleration in the
spinning neutron star~\cite{Blasi:xm}, resulting in an $E^{-1}$
spectrum. Softening of the spectrum to $E^{-2}$ ensues through
gravitational wave losses during spindown~\cite{Arons:2002yj}.
Following this scheme, we assume that some of the nuclei are
captured in the dense region of the source, attaining sufficient
diffusion in milliGauss magnetic fields. The resulting time delay
of several thousand years\cite{Bednarek:2001av} produces a
further steepening of the  injection power law spectrum. Note that
once diffusion has been established, additional Rayleigh steps in
the Galactic magnetic field do not change the spectral index
significantly. In their random traversal of the OB association,
the nuclei  undergo photodisintegration on far infrared thermal
photons  populating molecular clouds with temperatures of 15-100
K~\cite{Wilson:1997ud}. Taking an average photodisintegration
cross section of 40~mb, we find an interaction time between 4 and
1300~yr, allowing sufficient neutron production to explain the
anisotropy.

To incorporate the preceding discussion in our work,
we take in what follows a single power law neutron
spectrum with a spectral index of 3.1, representing an average
over the PeV-EeV energy region. Specifically,
$dF_n/dE_n = C\, E_n^{-3.1},$ with the normalization constant fixed
near an EeV to the observed excess. The constant $C$ is
determined by integrating $dF_n/dE_n$ over a bin $(E_1, E_2)$
with the result
\begin{equation}
F_n = \int_{E_1}^{E_2} C\,\, E_n^{-3.1} = 0.95\, \,C\,\,
\overline E_n^{-2.1}\,\,
{\rm sinh} \left[1.1\,\, \Delta \right]\,,
\end{equation}
where $\overline E_n = \sqrt{E_1\,E_2} \approx  10^{9.2}$~GeV,
$\Delta = \ln (E_2/E_1) \approx 1.38,$ and
$F_n \approx 9$~km$^{-2}$ yr$^{-1}$~\cite{Teshimaicrc}.

The typical energy for the antineutrino in the lab is that
of the decaying neutron times $Q/m_n\sim 10^{-3}$.
Thus, the stability of neutrons at $\gsim$~EeV implies a PeV upper limit
for the produced antineutrinos.
The increasing abundance of neutrons below an EeV in turn implies
an increasing neutrino flux as energies move below a PeV.

Nuclei with Lorentz factor $\sim 10^6$ are synthesized in all supernovae.
Hadronic interactions with the HII population
(density 30~cm$^{-3}$~\cite{Butt:2003xc}) and photodisintegration from
ultraviolet photons emitted from OB stars results in a flux
of PeV neutrons. From the measured Lyman emission of the O
stars~\cite{Knodlseder:2000vq}, the O$\div$B
temperature, and luminosity
characteristics of the B stars~\cite{Hanson:2003mf}, we obtain a photon number
density in the ultraviolet
of $\sim$ 230 cm$^{-3}$ for the core region $\sim$ 10 pc.
This implies a collision time
of about 0.1~Myr, comparable to the time scale for hadronic interactions.
However, in contrast to hadronic interactions,
significant photodisintegration occurs
in the outer 20~pc of the source. The photon density is
reduced to $\agt 25$ cm$^{-3}$, which lengthens the reaction time to
$\sim$ 1 Myr. The diffusion time ($\sim 1.2$ Myr) is a bit smaller
than the age of the
cluster $\sim$ 2.5 Myr~\cite{Knodlseder:2001eb}, and somewhat higher
than the reaction time.
This is  sufficient to permit over 90\% efficiency
for photodisintegration over the lifetime of the source.
The effective volume for
photodisintegration is then about a factor of 27 larger than for hadronic
interactions, and the net result of all these considerations is that the
PeV neutron population is about an order of magnitude greater than TeV charged
pions resulting from hadronic collisions~\cite{Knapp:tv}.

Performing the $\cth$ -integration in Eq.~\rf{nuflux}
over the delta-function constraint leads to
\begin{eqnarray}
\label{nuflux2}
\frac{d\Fnu}{d\Enu}(\Enu) &  = & \frac{m_n}{2}\,
\int_{\Enmin} \frac{d\En}{\En}\,\frac{d\Fn}{d\En}(\En)
\left(1-e^{-\frac{D\,m_n}{\En\,\tbar}}\right)\nonumber \\
 & \times &
\int_{\epsmin}^Q \frac{d\Enubar}{\Enubar}\,
\frac{dP}{d\Enubar}(\Enubar)\,,
\end{eqnarray}
with $\epsmin=\frac{\Enu\,m_n}{2\En},$ and $\Enmin=\frac{\Enu\,m_n}{2Q}$.
An approximate answer is available.
Setting the beta-decay neutrino energy $\Enubar$ equal to its mean value
$\equiv \eps_0$, we have
$\frac{dP}{d\Enubar}(\Enubar)=\delta(\Enubar-\eps_0)$.\footnote{The delta-function
in the neutron frame gives rise to a flat
spectrum for the neutrino energy in the lab for fixed neutron
lab-energy $\En=\Gamma_n\,m_n$:
\[
\frac{dP}{d\Enu}=\int^1_{-1}\frac{d\cth}{2}\,
\left(\frac{d\Enubar}{d\Enu}\right)\,
\left(\frac{dP}{d\Enubar}\right) = \frac{1}{2\,\Gamma_n\,\eps_0 }
\,,
\]
with $0\le \Enu \le 2\,\Gamma_n\,\eps_0$.}
When the delta-function is substituted into Eq.~(\ref{nuflux2}),
one gets
\begin{equation}
\frac{d\Fnu}{d\Enu}(\Enu)=\frac{m_n}{2\,\eps_0}
\int_{\frac{m_n\,\Enu}{2\,\eps_0}} \frac{d\En}{\En}\,
    \frac{d\Fn}{d\En}(\En)
    \left(1-e^{-\frac{D\,m_n}{\En\,\tbar}}\right)\,.
\label{nuflux3}
\end{equation}

Further approximation is available.
Treating the neutron decay factor, $1-\exp(\cdots),$
as a step function $\Theta(\Enmax-\En)$ at some energy
$\Enmax\sim{\cal O}(\frac{D\,m_n}{\tbar})
=(\frac{D}{10{\rm kpc}})$~EeV,
i.e. the neutron is unstable for $\En<\Enmax$ and
stable for $\En>\Enmax$,
one obtains from Eq.~(\ref{nuflux3})
\begin{equation}
\label{nuflux4}
\frac{d\Fnu}{d\Enu}(\Enu)=\frac{m_n}{2\,\eps_0}
\int^{\Enmax}_{\frac{m_n\,\Enu}{2\,\eps_0}} \frac{d\En}{\En}\,
    \frac{d\Fn}{d\En}(\En)\,.
\end{equation}
We have found that for $E_n^{\rm max} = 3$~EeV, Eqs.~(\ref{nuflux3}) and
(\ref{nuflux4}) differ by less than 1\% for $\Enu$ as high as 1 PeV.

A direct $\nuebar$ event in IceCube
will make a showering event, which, even if seen,
provides little angular resolution.
In the energy region below 1~PeV, IceCube will resolve directionality
only for $\nu_\mu$ and  $\bar \nu_{\mu}.$
Fortunately, neutrino oscillations rescue the signal.
Since the distance to the Cygnus region greatly exceeds the
$\nuebar$ oscillation length
$\lambda_{\rm osc} \sim 10^{-2}\frac{E_{\bar \nu}}{\rm PeV}$~parsecs
(taking the solar oscillation scale
$\delta m^2 \sim 10^{-5}{\rm eV}^2$),
the antineutrinos decohere in transit.
The arriving antineutrinos are distributed over flavors,
with the muon antineutrino flux $F_{\bar \nu_\mu}$
given by the factor
${1 \over 4}\, \sin^2 (2\,\theta_\odot) \simeq 0.20$
times the original $F_{\bar \nu_e}$ flux.
The $\bar \nu_\tau$ flux is the same,
and the $\bar \nu_e$ flux is 0.6 times the original flux.
Here we have utilized for the solar mixing angle
the most recent SNO result
$\theta_\odot \simeq 32.5^\circ$ \cite{SNOnew},
along with maximal mixing for atmospheric $\nu_\mu$-$\nu_\tau$
neutrinos and a negligible $\nu_e$ component in
the third neutrino eigenstate.

\begin{figure}
\begin{center}
\includegraphics[height=8.cm]{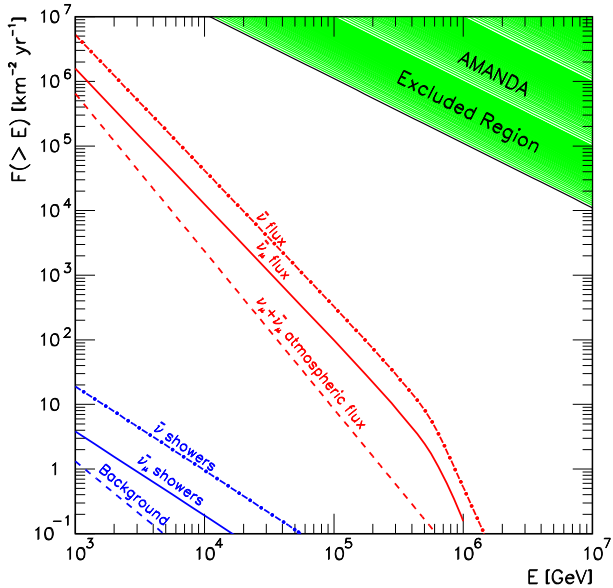}
\caption{Integrated flux of $\bar \nu_\mu$ (solid) and $\bar \nu_\mu + \bar \nu_e + \bar \nu_\tau$ (dashed-dotted)  predicted to arrive
at Earth from the direction of the Cygnus region.
Also shown  is the integrated $\nu_\mu + \bar\nu_\mu$ atmospheric
flux for an angular bin of $1^\circ \times 1^\circ.$  The shaded band
indicates the region excluded by the AMANDA experiment~\cite{Ahrens:2003pv}.
The expected number of showers $\bar \nu_\mu$ (solid) and   $\bar \nu_\mu + \bar \nu_e + \bar \nu_\tau$ (dashed-dotted)
to be detected (say in IceCube) are plotted on the bottom-left.
The expected background for the same angular bin is indicated by the dashed
line.}
\label{fig}
\end{center}
\end{figure}

The integral neutrino flux
$\Fnu(>\Enu)\equiv\int_{\Enu} d\Enu\,\frac{d\Fnu}{d\Enu}$
is particularly useful
for experiments having a neutrino detection-efficiency that is
independent of neutrino energy, or nearly so.
IceCube is an example of such an experiment.
Our calculated integral flux is shown in Fig.~\ref{fig}.
As mentioned above, the nuclear photodisintegration threshold leads to
an infrared cutoff on the primary neutron energy at the source,
which in turn leads to a low energy cutoff $\sim$~TeV on the integral flux.

We now  estimate the signal-to-noise ratio at IceCube. The angular
resolution of the experiment $\approx 0.7^\circ$ allows a search window of
$1^\circ \times 1^\circ$~\cite{Ahrens:2002dv}.
We begin with the ``noise''.
The event rate of the
atmospheric $\nu$-background that will be detected in the search bin
($\Delta \Omega_{1^\circ \times 1^\circ} \approx 3 \times 10^{-4}$~sr)
is given by
\begin{equation}
\left. \frac{dN}{dt}\right|_{{\rm background}} = A_{\rm eff}\,
\int dE
\,J_{\nu + \bar \nu}(E)\,\, p (E)
\,\,\Delta \Omega_{1^\circ \times 1^\circ}\,, \label{background}
\end{equation}
where $A_{\rm eff}$ is the effective area of the detector,
$J_{\nu + \bar \nu} (E)$  is the
$\nu_\mu + \bar \nu_\mu$ atmospheric flux in the direction of the
Cygnus region (about $40^\circ$
below the horizon)~\cite{Lipari:hd}, and $p (E) \approx 1.3 \times 10^{-6} \,(E/{\rm TeV})^{0.8}$
denotes the probability (generic to ice/water detectors)
that a $\nu$ (or $\bar \nu$) with energy $E$ on a
trajectory through the detector produces a signal~\cite{Gaisser:1994yf}.
For a year of running at IceCube and $E_{\bar \nu}^{\rm min} = 1$~TeV,
from Eq.~(\ref{background}) one obtains a background of 1.5 events.
Existing limits on the $\gamma$--ray flux from the Cygnus 
region~\cite{Borione:1996jw} provide an upper limit on  neutrino 
fluxes generated via $\pi^\pm$ decay at the source. This limit is below the atmospheric background in 
the region of interest. Poisson statistics then imply that a
signal $\ge~3.7$ events is significant at the 95\%
CL~\cite{Feldman:1997qc}.
The number of $\bar \nu$ showers in the signal,
for energies above $E_{\bar \nu}^{\rm min}$,
is given by
\begin{equation}
\label{signal}
\left. \frac{dN}{dt}\right|_{\rm signal}  =  A_{\rm eff}
\,\int_{E_{\bar \nu}^{\rm min}}
d\Enu\,\frac{d\Fnu}{d\Enu}(\Enu)\,\,p(\Enu)\, .
\end{equation}
For a year of running at IceCube, one expects 20 neutrino
showers (all flavors) with energies $\ge$ 1~TeV,
of which 4 $\bar \nu_\mu$ events will cluster within $1^\circ$
of the source direction, comfortably above the stated CL.
Neutrino flux at 1~TeV may also originate in the decay of 1 PeV 
neutrons from sources whose spectrum cuts off at that energy, 
and hence are not subject to normalization by the anisotropy. 
Thus our estimate may be regarded as very conservative.

IceCube is not sensitive to TeV neutrinos from the GC,
as these are above the IceCube horizon,
where atmospheric muons will dominate over any signal.
However, other kilometer-scale neutrino detectors, such as those planned
for the Mediterranean Sea, may see the GC flux.\footnote{The model depends
on dilation of the neutron lifetime due to standard
special relativity (SR).
For neutrons from the GC, SR predicts stability at $E_n \agt 1$~EeV
and decay at lower energies.
Thus, if the model is validated, say through measurement of the
predicted neutrino flux,
then SR is validated to an unprecedented boost factor of $\Gamma \approx 10^9$,
many orders of magnitude beyond present limits.} Additionally,
Southern Auger
should see the cosmic ray excess in the direction of the GC~\cite{Clay:2003pv}
and Northern Auger should be sensitive to the Cygnus region.

We conclude that in a few years of observation, IceCube will
attain  $5\sigma$ sensitivity for discovery of the
Fe$ \rightarrow n \rightarrow \nuebar\rightarrow \bar \nu_\mu$ cosmic
beam, providing the ``smoking ice'' for the GP neutron hypothesis.

\section*{Acknowledgments}

We thank John Beacom, Jim Cronin, Concha Gonzalez-Garcia, Todor Stanev, and
Alan Watson for valuable discussions.
This work has been partially supported by NSF, DoE, NASA, and the Wisconsin
Alumni Research Foundation.

\end{document}